\documentclass[apj]{emulateapj}
\usepackage{color}
\usepackage{epsf}
\usepackage{ulem}
\usepackage{graphicx}
\newcommand{\Ok}{\Omega_K}
\newcommand{\DA}{D\!_A(z)}
\newcommand{\hz}{H(z)}

\newcommand{\Vsur}{V_{\rm sur}}

\newcommand{\Oh}{\Omega_m h^2}
\newcommand{\Obhh}{\Omega_b h^2}

\newcommand{\hMpc}{h^{-1}{\rm\;Mpc}}

\newcommand{\trihGpc}{h^{-3}{\rm\;Gpc^3}}
\newcommand{\itrihMpc}{h^{3}{\rm\;Mpc^{-3}}}
\newcommand{\ihMpc}{h{\rm\;Mpc^{-1}}}
\newcommand{\kmax}{k_{\rm max}}

\newcommand{\nPt}{nP_{0.2}}
\newcommand{\sig}{\sigma}

\newcommand{\Df}{\Delta f}
\newcommand{\kb}{k_B}
\newcommand{\Tsky}{\bar{T}_{\rm sky}}
\newcommand{\Ta}{\bar{T}_a}
\newcommand{\Tsig}{\bar{T}_{\rm sig}}
\newcommand{\tint}{t_{\rm int}}
\newcommand{\Poh}{P_{\rm HI}}
\newcommand{\OHI}{\Omega_{\rm HI}}

\newcommand{\Pshot}{P_{\rm shot}}
\newcommand{\Nfeed}{N_{\rm feed}}
\newcommand{\Nyear}{N_{\rm year}}
\newcommand{\Asur}{A_{\rm survey}}
\newcommand{\Lcyl}{L_{\rm cyl}}
\newcommand{\Wcyl}{W_{\rm cyl}}

\newcommand{\knyq}{k_{\rm Nyq}}
\newcommand{\hdel}{\hat{\delta}}

\begin{document}
\title{A ground-based 21cm Baryon acoustic oscillation survey}

\author{
Hee-Jong Seo\altaffilmark{1}, Scott Dodelson\altaffilmark{1,2}, John Marriner\altaffilmark{1}, Dave Mcginnis\altaffilmark{1}, Albert Stebbins\altaffilmark{1}, Chris Stoughton\altaffilmark{1}, Alberto Vallinotto\altaffilmark{1}}

\begin{abstract}
Baryon acoustic oscillations (BAO) provide a robust standard ruler with which to measure the acceleration of the Universe. 
The BAO feature has so far been detected in optical galaxy surveys. Intensity mapping of neutral hydrogen emission with a ground-based radio telescope provides another promising window for measuring BAO at redshifts of order unity for relatively low cost. While the cylindrical radio telescope (CRT) proposed for these measurements will have excellent redshift resolution, it will suffer from poor angular resolution (arcminutes at best). We investigate the effect of angular resolution on the standard ruler test with BAO, using the Dark Energy Task Force Figure of Merit as a benchmark. We then extend the analysis to include variations in the parameters characterizing the telescope and the underlying physics. Finally, we optimize the survey parameters (holding total cost fixed) and present an example of a CRT BAO survey that is competitive with Stage III dark energy experiments. 
The tools developed here form the backbone of a publicly available code that can be used to obtain estimates of cost and Figure of Merit for any set of survey parameters.
\end{abstract}
\altaffiltext{1}{Center for Particle Astrophysics, Fermi National Accelerator Laboratory, P.O. Box 500, Batavia, IL 60510-5011; sheejong@fnal.gov}
\altaffiltext{2}{Department of Astronomy \& Astrophysics, The University of Chicago, Chicago, IL 60637-1433}

\section{Introduction}
A standard ruler test with Baryon acoustic oscillations (BAO) is considered the most robust and systematics-free method to probe the dark energy equation of state \citep{Albrecht:2006um}. The sound waves which propagated through a mixture of photons and baryons in the early Universe left a distinct oscillatory signature in the Cosmic Microwave Background (CMB) and the large scale structure of matter (or galaxies) with a characteristic scale set by the sound horizon at the epoch of recombination \citep{Peebles70,Holtzman89,Hu96,EH98}. Measurements of the CMB provide this sound horizon scale, and we can then use the BAO in low-redshift clustering as a standard ruler to measure the distances to various redshifts, and therefore the equation of state of dark energy \citep{HW96,EHT98,Eisen03,Blake03,Linder03,Hu03,SE03}. 

While detections of BAO so far have used the optical band of the electromagnetic spectrum, using spectroscopic or photometric techniques, there has been growing interest in the feasibility of using 21cm emission from neutral hydrogen \citep{WL08,LW08,WL09,Visbal09} to detect BAO at high \citep{Mao:2008ug,WLal08} and medium  \citep{Chang:2007xk,Ansari:2008yw} redshifts. Observing BAO from intensity mapping of 21cm emission, especially at medium redshift near $z\sim 1$, has several advantageous features. The BAO is a relatively weak feature on large scales, and therefore very large survey volumes are needed. A ground-based cylindrical radio telescope (CRT) with a large field of view can easily cover most of the sky in a relatively short time. Second, the electronics required for frequencies near $\nu\sim 1{\rm GHz}$ are cheap and easy to build. Digital electronics with precise timing offer high precision (better than ppm) frequency and, hence, redshift measurements. Third, radio surveys rely on different tracers of large scale structure (neutral hydrogen) than do optical survey (luminous galaxies). Seeing the signal in two different sets of mass tracers would be compelling evidence for its robustness. 

A disadvantage of radio waves is their longer wavelengths, making it more difficult to obtain high angular resolution.
For example, a 100m by 100m telescope will have an angular resolution of $(\lambda/L) \sim 15$ arcminutes at $z\sim 1$. Radio observations of cosmological hydrogen are severely contaminated by foreground sources, primarily Galactic synchrotron radiation. We will not consider the problem of foreground removal here, but will show that foregrounds are smooth at the BAO scale and then assume that they can be eliminated to the accuracy of the statistical errors. 

In this paper, we consider a ground-based 21cm intensity mapping with a CRT targeting a redshift range of 0.2-2, motivated by the study of \citet{Chang:2007xk}. In \S~\ref{sec:StoN}, we derive the signal-to-noise of the power spectrum for a given set of radio telescope configuration parameters, discuss galactic shot noise and the effect of foregrounds, and explain the assumptions used in our method of Fisher matrix projections. We model the effects of angular resolution with a window function that approximates the effects of a typical CRT telescope response function on the power spectrum.  Using both Fisher matrix analysis and Monte Carlos, we study the issue of angular resolution in detail in \S~\ref{sec:antheta}, focusing on the impact of resolution on the determination of dark energy parameters. 
Then in \S~\ref{sec:FoM}, we consider a simple compact CRT telescope, and investigate the Dark Energy Task Force (DETF) Figure of Merit (FoM) \citep{Albrecht:2006um} as a function of telescope parameters. The main result of the paper is captured in Fig.~\ref{fig:figparam} where the FoM is plotted as a function of these parameters. 
In \S~\ref{sec:Max}, we optimize the telescope configuration parameters for the maximum FoM and show that an inexpensive CRT BAO survey can be competitive with Stage III dark energy experiments \citep[for optimazation studies in optical surveys, see][]{Parkinson07,Parkinson09}. In \S~\ref{sec:con} we summarize the results.

\section{Ingredients for Projections in a Radio telescope}\label{sec:StoN}

In this section, we derive the signal-to-noise ratio in the power spectrum for a given configuration of a cylindrical radio telescope. Once the signal-to-noise ratio per Fourier mode is determined, it is straightforward to propagate this to the errors on the acoustic scale and ultimately on dark energy parameters.

\subsection{Signal to Noise Ratio}

In a radio telescope, the measurement is power (i.e., Watts) received by the antenna, and we denote this with a lower case $p$ to distinguish it from the power spectrum $P(k)$ in wavenumber space. The average and the standard deviation of power received by an antenna per bandwidth $\Df$ due to the instrumental and sky noise are
\begin{eqnarray}
p_N&=&\kb (g\Tsky+\Ta)\Df,\label{eq:pN}\\
\sigma_{p_N}&=&p_N,
\end{eqnarray}
where $\kb$ is Boltzmann's constant, $g$ is the gain, $\Tsky$ is the average sky temperature (e.g., due to foregrounds), and $\Ta$ is the average antenna noise temperature or the amplifier noise temperature.  After $M$ such measurements, the uncertainty associated with $p_N$ per pixel for a compact array telescope is
\begin{eqnarray}
\sigma_{p_N}&=&\frac{\kb (g\Tsky+\Ta)\Df}{\sqrt{M}}=\frac{\kb (g\Tsky+\Ta)\Df}{\sqrt{\tint \Df}},\label{eq:spN}
\end{eqnarray}
where $\tint$ is the integration time per pixel per bandwidth.

Meanwhile, the BAO signal is the spatial temperature variation due to the clustering of neutral hydrogen, $\delta_{p_S}(\hat n,z)=p_S \delta_{\rm HI}(\hat n, z)$ where $\delta_{\rm HI}(\hat{n},z)$ is the fractional overdensity of neutral hydrogen at angular position $\hat{n}$ and at redshift $z$. The average power (again in, e.g., Watts) due to the 21cm line is
\begin{equation}
p_S=\kb g\Tsig \Df, \label{eq:pS}
\end{equation}
 with $\Tsig$ the average brightness temperature due to the 21cm line. This has been estimated \citep{Barkana07,Prit08,Chang:2007xk} to be

\begin{eqnarray}
\Tsig=188 \frac{x_{\rm HI}(z)\Omega_{\rm H,0} h (1+z)^2}{H(z)/H_0} mK,
\end{eqnarray}
where $x_{\rm HI}$ is the neutral hydrogen fraction at $z$, $\Omega_{\rm H,0}$ is the ratio of the hydrogen mass density to the critical density at $z=0$, and $H(z)$ is the Hubble parameter at $z$.

The measured power spectrum of the 21cm intensity $\hat{P}(k,\mu)$ will include both signal and noise: 
\begin{eqnarray}
\hat{P}(k,\mu)&=&p^2_S P_{\rm HI}(k,\mu) + V_R \sigma^2_{p_N},
\end{eqnarray}
where  $P_{\rm HI}(k,\mu)$ is the power (in units of $h^{-3}{\rm Mpc}^3$) due to the large scale structure of the neutral hydrogen (i.e., the signal) and $V_R$ is the volume of a pixel (also with units of $h^{-3}{\rm Mpc}^3$). We present a simple derivation of this equation in Appendix A. Since we are interested in structure on large scales, we assume that $\Poh(k,\mu)=b^2P(k,\mu)$ where $P(k,\mu)$ is the underlying matter power spectrum with linear theory redshift distortions \citep{Kaiser87}, $\mu$ is the cosine of the angle between the wavevector $\vec{k}$ and the line of sight, and $b$ is a constant bias factor.

Additionally, there will be a galactic shot noise contribution due to the discreteness of the HI sources with an effective number density $\bar{n}$. Adding this noise leads to our final expression for the signal to noise per Fourier mode:
\begin{equation}
\frac{S}{N}=\frac{\Poh}{\Poh+\left[\frac{(g\Tsky+\Ta)}{g\Tsig\sqrt{\tint \Df}}\right]^2 V_R+\frac{1}{\bar{n}}}, 
\end{equation}
where ${(\Tsky+\Ta/g)}\over{\sqrt{\tint \Df}}$ is the sensitivity per pixel.
As will be explained in \S~\ref{subsec:shot}, we approximate $\Pshot=1/\bar{n}=100 h^{-3}$ Mpc$^3$. Assuming our fiducial CRT configuration as explained in \S~\ref{subsec:fidCRT}, $\Poh(k,\mu=0) \sim  900 h^{-3}$ Mpc$^3$ and $\left[\frac{(g\Tsky+\Ta)}{g\Tsig\sqrt{\tint \Df}}\right]^2 V_R \sim 1800h^{-3}$ Mpc$^3$ at $z\sim 1$. 

 For a power spectrum averaged over a wavenumber range of width $dk$, the signal to noise is increased by the square root of the number of modes, so
\begin{eqnarray}
\frac{S}{N}&=&\sqrt{\frac{2\pi k^2dk d\mu\Vsur}{2(2\pi)^3}}\frac{\Poh(k,\mu)}{\Poh(k,\mu)+\left[\frac{(g\Tsky+\Ta)}{g\Tsig \sqrt{\tint \Df}}\right]^2 V_R+\frac{1}{\bar{n}}}, \nonumber \\
&&
\end{eqnarray}
where $\Vsur$ is the total volume of the survey.

We modify this equation further by including a window function $\hat{W}$ that approximates the effect of an instrument response function.

\begin{eqnarray}
\frac{S}{N}&=&\sqrt{\frac{2\pi k^2dk d\mu\Vsur}{2(2\pi)^3}}\frac{\Poh(k,\mu)\hat{W}^2}{\Poh(k,\mu)\hat{W}^2+\left[\frac{(g\Tsky+\Ta)}{g\Tsig \sqrt{\tint \Df}}\right]^2 V_R+\frac{1}{\bar{n}}}. \nonumber \\
&&\label{eq:StoN}
\end{eqnarray}

We discuss details of $\hat{W}$ in \S~\ref{subsec:Fisher}.

\subsection{Telescope parameters for our fiducial compact array CRT}\label{subsec:fidCRT}

\begin{figure}
\includegraphics[angle=-90,width=8cm,totalheight=5cm]{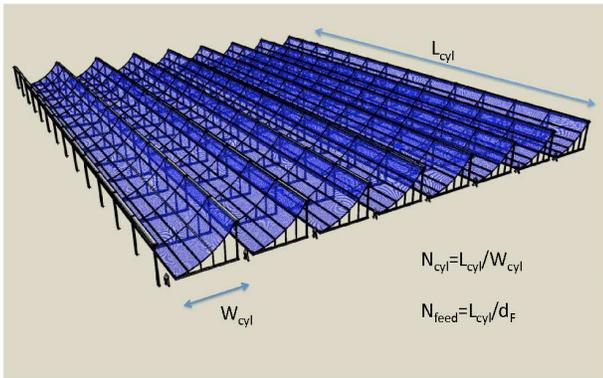}
\caption{General configuration of a compact CRT array. $\Lcyl$ is the length of the cylinder, $\Wcyl$ is the width of the cylinder. We assume the cylinder is square and compact so $\Wcyl$ times the number of cylinders is set equal to $\Lcyl$. Also the number of antenna feeds along a single cylinder is equal to the length of the cylinder divided by $d_F$, the spacing of the receivers.}\label{fig:CRT}
\end{figure}

\begin{figure}
\epsscale{1.2}
\plotone{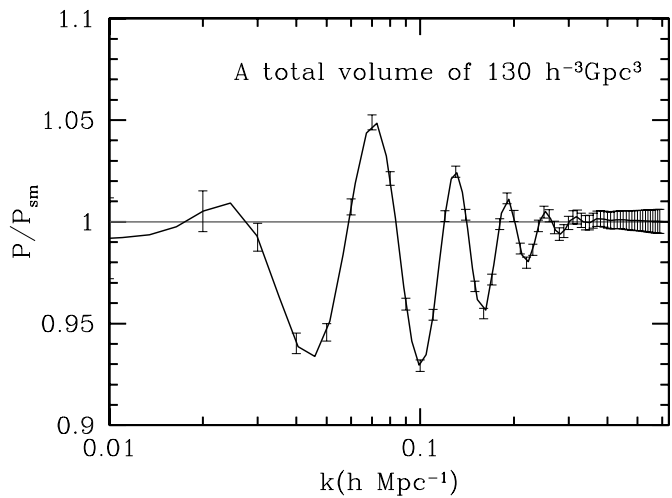}
\caption{Expected errors on the power spectrum for our fiducial survey in Table \ref{tab:fid}. We assume a power spectrum at $z=1$ in real space, after properly accounting for the nonlinear degradation on BAO, and the instrumental noise at $z=1$ while assuming the entire survey volume. A wavenumber bin width of $0.01\ihMpc$ is assumed. }\label{fig:fid}
\end{figure}

The input parameters required to achieve a given signal to noise (Eq. \ref{eq:StoN}) depend on the configuration of the telescope. The important parameters are the length of the cylinder $\Lcyl$, the width of the cylinder $\Wcyl$,  the spacing of receivers $d_F$ (or the number of receivers $\Nfeed=\Lcyl/d_F$), and the total survey time $\Nyear$. We consider a rectangular compact array  so that $N_{\rm cyl}=\Lcyl/\Wcyl$. Fig. \ref{fig:CRT} shows the general configuration for the compact array. For this configuration,

\begin{eqnarray}
&&\mbox{\it Resolution of the beam width, }\Delta \theta_{\rm res} \approx \frac{\lambda}{\Lcyl},\\
&&\mbox{\it Area of the survey, }\Asur= \int_0^{2\pi}d\phi \int_{\theta_{\rm lat}-\Delta \theta/2}^{\theta_{\rm lat}+\Delta \theta/2} cos(\theta) d\theta \nonumber \\
&&\stackrel{\theta_{\rm lat}=0} {\Longrightarrow} 4\pi sin\left( \frac{\Delta\theta}{2}\right)  \approx 2\pi\frac{\lambda}{d_F}=2\pi{\lambda}\frac{\Nfeed}{\Lcyl},\label{eq:Area}\\
&&\mbox{\it Integration time per pixel, } \tint \approx \Nyear D_f \frac{1}{2\pi}\frac{\lambda}{\Wcyl}.\nonumber\\
&&
\end{eqnarray}

Above and throughout, we have made the small angle approximation, assumed critical sampling, and neglected side lobes. We also assume a CRT telescope at the equator as our fiducial configuration for simplicity, which, according to equation (\ref{eq:Area}), affords all steradian sky coverage in the limit of half-wavelength feed spacing resulting in coverage to the horizon. We consider the perhaps more likely case of a telescope at a latitude of $\sim 30$ degrees in section \S~\ref{sec:Max}.
The angular resolution and the frequency resolution $\Delta f$ determine the pixel size, and therefore $V_R$, and Asurvey determines Vsurvey. Note that the effect of $\Delta f$, i.e., the resolution along the line of sight cancels out, as it appears both in the sensitivity per pixel and $V_R$. We nevertheless assume that the frequency resolution is good enough (i.e., better than 1MHz, which can be easily achievable in this survey) that we can ignore the instrument response function in this direction (see \S~\ref{subsec:Fisher}).
We fix the remaining parameters such as the antenna temperature, sky temperature, gain, and the observing duty factor to be $\Ta=50K$, $\Tsky=10K$, $g=0.8$, and $D_f=0.5$. We make a conservative assumption for the neutral hydrogen fraction: $x_{\rm HI}(z)\Omega_{\rm H,0} = 0.00037$ \citep{Zwaan:2005cz,igm:phw05,igm:rtn06}.

\begin{center}
\begin{deluxetable}{cccccc}
\tablewidth{0pt}
\tabletypesize{\footnotesize}
\tablecaption{\label{tab:fid} Fiducial CRT configuration.}
\startdata \hline\hline
&  Low redshift  & High redshift \\ \hline
Parameters & $0.66 <z<1.24$  & $1.22 < z < 2.11$ \\ \hline 
Length of Cylinder, $\Lcyl$ (m)&	99.8	&	142.8 \\
Feed spacing, $d_F$ (m)		&	0.39	&	0.558 \\
Width of Cylinder, $\Wcyl$ (m)  & 	14.3	&	14.3  \\	 
Duty factor, $D_f$    	&	0.5	&	0.5	\\
$\Nyear$ (years)	&	1.40	&	0.87	\\
$x_{\rm HI}\Omega_{\rm H,0}$			& 	0.00037	&	0.00037	\\
bias			&	1.0	&	1.0 \\
Sky temperature, $\Tsky$ (K)		& 	10	&	10 \\
Antenna temperature, $\Ta$  (K)		& 	50	&	50	\\
gain, g			& 	0.8	& 0.8	\\
$\Pshot$		&	100.0	& 100.0	
\enddata
\tablecomments{For technical reasons, we assumed that the redshift range is covered by two distinct configurations of the telescope.}
\end{deluxetable}
\end{center}

Table \ref{tab:fid} lists the telescope parameters for a fiducial CRT survey. For technical reasons \footnote{It is difficult to design an array with large fractional bandwidths ($> 50\%$). By dividing the design into two configurations, the fractional bandwidth of our fiducial CRT falls into a reasonable range of about 30\%.}, the redshift range will likely be covered by two distinct configurations of the telescope: in our fiducial survey, the first configuration covers $0.60 < z <1.24$ and the second configuration covers $1.22 < z < 2.11$. These will be combined to obtain estimates of the power spectrum in a series of redshift bins of size $\Delta z =0.1$. The signal to noise on the power spectrum will depend on the redshift bin under consideration. Figure \ref{fig:fid} shows the power spectrum at $z=1$ with associated errors derived from equation (\ref{eq:StoN}) for the fiducial survey. To give a sense of the constraining power of 21 cm surveys, the error bars in this figure assume that the total volume of the survey is concentrated at $z=1$.

\subsection{Galactic shot noise estimation}\label{subsec:shot}
In 21cm intensity mapping we are only able to probe the mean 21cm emission in fairly large volumes of space.  While intensity mapping has the advantage that it includes the emission from all the galaxies, no matter how dim, it has the disadvantage that, due to the luminosity weighting, the random sampling of the density field is dominated by a relatively small number of the bright galaxies.  This sampling noise is known as galactic shot noise, and is often calculated under assumption that the galaxy position and luminosity is randomly selected with probability proportional to $1+b\,\delta_{\rm m}$ where $b$ is the bias and $\delta_{\rm m}$ is the mass overdensity.  Under these assumptions the shot noise adds a scale independent $1/\bar{n}(z)$ term to the observed overdensity power. Using the Schechter function fit to the HIMF (eutral hydrogen mass function) by \citet{Zwaan:2005cz} we find $\bar{n}={\theta_*\over{\rm ln}10}\,{\Gamma(2+\alpha)\over2+\alpha}=0.01 \itrihMpc$ which we use in our calculations without evolution.  The evolution of $\bar{n}(z)$ is uncertain but shot noise is not the dominant source of noise.

\subsection{Foregrounds}\label{subsec:fore}
At the frequencies relevant for this study 21cm emission is far from the dominant source of emission.  Synchrotron (plus a smaller contribution of free-free) emission will dominate the 21 cm signal by a factor $\sim 10^4$.   This includes emission from our Galaxy and from extra-Galactic sources.  However as noted by a number of authors \citep{DiMatteo:2001gg,Zaldarriaga:2003du,Wang:2005zj,Morales:2005qk,Chang:2007xk} synchrotron (and free-free) emission will have a very smooth spectrum unlike the BAO wiggles and one can therefore distinguish the two.  In fact from the intensity maps one can extract ``modes'' which will be largely uncontaminated by foregrounds.  Furthermore these uncontaminated modes are the ones which are relevant for BAO analysis, and this is why one can calculate the Fisher information without taking account of the foregrounds.  This result is suggested in Figure 3 of \citet{Chang:2007xk}  which shows a significant residual foreground contamination for $k<0.04\ihMpc$, {\it i.e.} on scales larger than the BAO scale. Even so their model for foreground residuals is extremely pessimistic and one can reduce these significantly more than they have assumed. To further justify this statement we give an outline of how we have set model-independent lower limits for just how smooth the synchrotron spectrum must be and also present numerical results.  More specifics can be found in \citet{Stebbins:2010}. We decompose the observed intensity pattern into modes:
\begin{equation}
\hat{\delta}_I(\vec{k})=\int d^2{\hat n} \int d\nu\,m_{\vec{k}}(\hat{n},\nu)\,I_\nu({\hat n})\ .\end{equation}
For the BAO analysis one will want to use modes which are localized in the spatial wavenumber, $k$, describing the spatial
 pattern of 21cm emission.  One will furthermore want these modes localized in angular wavenumber, $l$, allowing one to distinguish modes whose variation is primarily radial or angular.

\begin{figure}
\epsscale{1.2}
\plotone{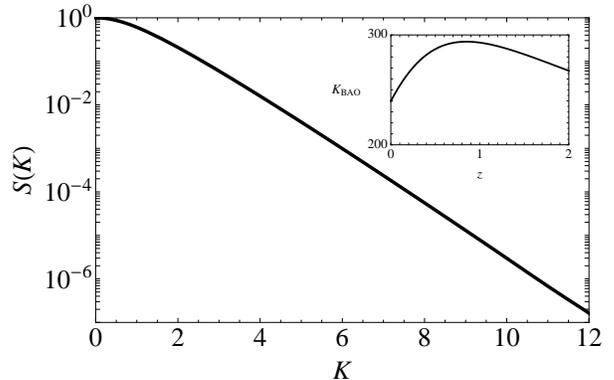}
\caption{Plotted is the suppression factor, $S(K)$, which is the ratio of the maximum amplitude in spectral wiggles in synchrotron intensity in units of the mean intensity.  It is a function of the logarithmic wavenumber $K$.  In the inset is plotted the logarithmic wavenumber corresponding to the 1st BAO peak as a function of redshift.  Extrapolating to get $S(K_{\rm BAO})<10^{-100}$.  These functions are computed in \citet{Stebbins:2010}.}\label{fig:suppression}
\end{figure}

The synchrotron intensity is given by
\begin{equation}
 I_\nu^{\rm sync}({\hat n})=\int_0^\infty {d N_{\rm e}\over d\epsilon} W({\nu \over \epsilon})\,d\epsilon
\end{equation}
where $\epsilon=E_{\rm e}^2\,|{\hat n}\times B|$, $E_{\rm e}$ is the electron energy, $B$ the magnetic field, and ${d N_{\rm e}\over d\epsilon}$ gives the distribution of $\epsilon$ along line-of-sight $\hat n$. $W$ is a function that describes synchrotron emission for given electron \citep{Ryb}. This convolution means that small-scale wiggles in the ${d N_{\rm e}\over d\epsilon}$ distribution function show up as wiggles in the spectrum of  $I_\nu^{\rm sync}$.  Since $W\ge 0$ and is smooth one finds that, relative to the mean, wiggles in $I_\nu^{\rm sync}$ are suppressed relative to wiggles in ${d N_{\rm e}\over d\epsilon}$, and the suppression is greater as one looks at higher frequency wiggles.  Furthermore since ${d N_{\rm e}\over d\epsilon}\ge 0$ one finds that the amplitude of wiggles in ${d N_{\rm e}\over d\epsilon}$ can be no bigger than the mean.  Given these constraints it is impossible for the amplitude of wiggles in $I_\nu^{\rm sync}$ to be larger than a suppression factor times the mean $I_\nu^{\rm sync}$.  At the frequencies of interest the mean $I_\nu^{\rm sync}$ corresponding to a brightness temperature of a few Kelvin so the physical bound on wiggles is the suppression factor times a few kelvin.  The suppression factor will depend on the physical scale of the wiggles and is extremely small for wiggles which could masquerade as BAO.

More specifically if one Fourier transforms the intensity in ${\rm ln}\,\nu$:  $\tilde{I}(K)=\int_{-\infty}^\infty I_\nu^{\rm sync} e^{-i K\,{\rm ln}\,\nu}d\nu$ then $|\tilde{I}(K)|\le S(K)\tilde{I}(0)$ where $S(K)\ge 0$ is the suppression factor which is plotted in Figure \ref{fig:suppression} for small $K$.  $S(K)$ is given by the logarithmic Fourier transform of $W$.  At the large $K$ relevant to BAO the $S(K)$ is so small that it is difficult to determine numerically, but since the Fourier transform of $C^\infty$ functions, such as $W$, should fall off exponentially one can extrapolate the $S(K_{\rm BAO})<10^{-100}$ where $K_{\rm BAO}$ is the logarithmic wavenumber corresponding to the BAO scale.  If one chooses modes which limit the amount of small-$K$ contamination one can greatly suppress the amount of possible synchrotron contamination.  One can apply exactly the same argument to free-free emission, which depends on the distribution of gas temperatures in position or $\epsilon$, and obtain similar levels of suppression.   In practice suppression factors of a $10^{-100}$ are not achievable due to the contamination from small $K$, and this level is not really needed since this is much larger than many other sources of noise such as the photon shot noise.  However if one understands the beam pattern and frequency response of the telescope, one can expect to achieve suppression factors of $<10^{-4}$ for most modes on the BAO scale, which is more than enough to make foreground contamination a negligible source of noise.

The CRT will be sensitive not only to total intensity but also to linear polarization.  While at emission the synchrotron light is linearly polarized with a smooth spectrum, Faraday rotation can cause small scale wiggles in the spectrum of each component of linear polarization by the time it reaches the CRT.  This does not effect the total intensity.  To avoid possible contamination by Faraday rotation one should combine the radio signals so as to make an intensity map with as small a contamination linear polarization as possible.

In this paper, we assume that the foreground is properly subtracted, and consider only the noise (i.e, fluctuations) associated with it.

\subsection{Fisher matrix calculations and instrument response}\label{subsec:Fisher}
Once we have estimates of the noise relative to the signal for a given survey and instrumental parameters, as in equation (\ref{eq:StoN}), we can derive error estimates on the angular diameter distance $\DA$ and Hubble parameter $\hz$ at various redshift bins (of size $\Delta z = 0.1$) using the Fisher matrix formalism presented in \citet{Seo:2007ns}: we consider nonlinear degradation on BAO both due to nonlinear structure growth and redshift distortions at each redshift. We combine this with the Planck Fisher matrix from the DETF and derive errors on dark energy parameters and the Figure of Merit (FoM) \citep{Albrecht:2006um}. The FoM is defined as the inverse volume of the 95\% confidence ellipse in the space of $w_0$ and $w_a$. 
Following the convention of the DETF, our parameters include the matter density $\Oh$, baryon density $\Obhh$, dark energy fraction $\Omega_{\rm de}$, curvature fraction $\Ok$, the spectral tilt $n_s$, and the amplitude $A$, and two dark energy equation-of-state parameters: $w_0$ and $w_a$.  We choose our fiducial cosmology to be consistent with WMAP1 \citep{Spergel:2003cb}.

Recent results \citep{Eisenstein:2006nk,Seo:2008yx} have shown that when the signal to noise ratio per Fourier mode is of order two, we can partially undo the nonlinear degradation of the BAO signal. For our Fisher projection, we assume that we can conduct reconstruction and halve the nonlinear damping scale of the BAO only when $\Poh(k,\mu=0)/(P_N+\Pshot)$ at $k=0.2\ihMpc$ is larger than two. We hereafter denote $\Poh(k,\mu=0)/(P_N+\Pshot)$ at $k=0.2\ihMpc$ as $\nPt$. 
\\

While we do not attempt to include an exact instrument response function, we adopt a reasonable approximation that describes the general response function of CRT in our Fisher matrix analysis. The CRT cylinders are oriented in a North-South line so that the field of view has a narrow width in right ascension (because of the focussing of the cylinders) but a very broad coverage in declination. As a consequence the beams formed scan the sky at fixed declination and the largest beam separation defines a maximum spatial Fourier component (i.e., the Nyquist frequency $\knyq  =\pi/{\rm Resolution}$) that can be measured in declination. Higher modes are aliased onto lower modes but the alias effect is greatly suppressed because of the non-zero beam width. The rotation of the earth results in nearly continuous sampling in right ascension. The CRT therefore has some sensitivity to all modes beyond $\knyq$ but the sensitivity decreases around $\knyq$ according to a window function which arises from the non-zero beam size. We model the beam shape assuming a uniform illumination of the cylinder aperture and calculate the beams that can be synthesized from the cross-correlations between cylinders. In both directions angular resolution is therefore given by the overall dimension of the array. The resulting expression is cumbersome but can be adequately represented for our purposes by an exponential damping of a signal with a characteristic damping scale set by the Nyquist limit $\knyq$ for a given angular resolution. As the damping effectively suppresses signals from the modes beyond the Nyquist limit along the right ascension, we refer to the effects in both directions as Nyquist limit or Nyquist cutoff from now on. Such limitation does not apply to wavemodes along the frequency direction (i.e., the line of sight) where the 21cm survey can achieve an excellent redshift resolution. For example, a frequency resolution better than 1MHz will give a resolution along the line of sight direction better than $8\hMpc$ even at $z=2$. We therefore apply no Nyquist limit along this direction. We therefore introduce the following anisotropic window function $\hat{W}(\vec{k})$ and include in the signal to noise calculation in equation (\ref{eq:StoN}):

\begin{eqnarray}
\hat{W}(\vec{k}) &=& \exp \left[-1.5(k_{\rm x}/\kmax)^2\right]\Theta(\kmax-k_{\rm y}) \label{eq:window}
\end{eqnarray}
where $\Theta$ is a Heaviside step function, $k_{\rm x}$ is a wavevector component in the direction of right ascension, $k_{\rm y}$ is a component in the direction of declination, and $\kmax$ is set to be the Nyquist frequency $\knyq$ given the angular resolution. In a real survey, an instrument response will generate a window function with more complex features than described above. However, note that the effect of the instrument response on the shape of the power spectrum can be precisely predicted and therefore corrected for in a real survey. We use equation \ref{eq:StoN} and \ref{eq:window} as our default for \S~\ref{sec:FoM}.  We explain the effect of including this high $k$ cutoff in more detail in \S~\ref{sec:antheta}.

An interferometer that forms beams using only cross-correlations also has a low $k$ cut-off that depends on the instantaneous field of view along the drift scan direction (i.e., in right ascension) but only on the plane $k_{\rm Dec}(=k_y)=0$. We have not explicitly corrected for the reduced sensitivity at low $k$ due to the limited field of view, but the resulting effect on the total number of modes per $k$ is minimal for the array geometries we consider, so we can safely ignore this effect. We note, in addition, that the beams that are formed can generally contain signals from more than one direction.  In our analysis, note that we assume that the antenna pattern cuts off sufficiently quickly outside the region where the beams are formed so that the signal from side lobes are attenuated to a negligible level.

\section{Effect of angular resolution}\label{sec:antheta}

\begin{figure*}
\plotone{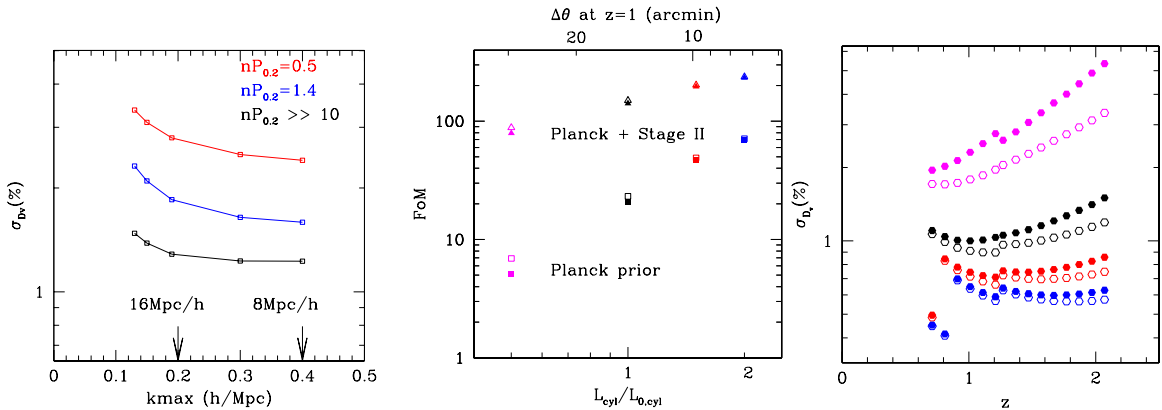}
\caption{The effect of introducing the Nyquist frequency limit $\kmax$ (i.e., $\hat{W}$ in eq. [\ref{eq:StoN}] and [\ref{eq:window}]) into the Fisher matrix calculation to account for a limited angular resolution. Left: Fisher matrix estimates of errors on the distance scale as a function of $\kmax$ for different cases of $\nPt$ assuming a BAO feature at $z=1$ in redshift space. The arrows point at $\kmax=\knyq$ for the resolution of $16\hMpc$ and $8\hMpc$. Middle: the difference in FoM between including (solid points) and ignoring the Nyquist frequency limit (open points) as a function of $\Lcyl$, i.e., as a function of an angular resolution for our fiducial CRT configuration. The squares are with the Planck prior and triangles are with the Planck + Stage II priors. Right: the corresponding errors on the distance scale $D_V$ at each redshift bin of the fiducial CRT. The solid points are with the Nyquist frequency cutoff and the open points are without the cutoff. The same colored points in the middle and the right panels correspond to the same value of $\Lcyl$. }\label{fig:kmax}
\end{figure*}

Unlike optical galaxy surveys where the angular resolution is of order arcseconds, the expected angular resolution of a CRT is a few arcminutes at best. Such a limited angular resolution might potentially impair the acoustic peak measurement and therefore distance measurements in various ways. The angular resolution of a CRT affects the pixel volume $V_R$ in equation (\ref{eq:StoN}), but also determines the range of the wavenumbers available due to the cutoff near the Nyquist frequency as modeled with the window function $\hat{W}$. While the former effect is obvious, the latter has not been investigated previously. Before we move to the DETF FoM for a CRT, we investigate the effects of this window function on the acoustic peak measurement and test using Fisher matrix analysis and Monte-Carlo simulations.

We focus on the following effects of angular resolution (i.e., the window function).
\\

$\bullet$ The increase in errors due to the wavenumber cutoff near the Nyquist frequency $\knyq$. We can test this analytically within the framework of Fisher matrix.

$\bullet$ Instability of the $\chi^2$ fitting due to the short wavenumber range available below $\knyq$. We test this using Monte-Carlo simulations.
\\

First, limited angular resolution will decrease S/N not only by increasing $V_R$ but also by limiting the wavenumbers available. Note that the Fisher matrix method from \citet{Seo:2007ns} assumes that the power spectrum up to a large enough wavenumber (i.e., $\kmax=0.5\ihMpc$) is available from observation, implicitly relying on an optical/IR survey. However, the resolution of radio surveys will likely limit $\knyq$, therefore $\kmax$, to be much below $0.5\ihMpc$.
Due to the degradation of BAO by Silk damping \citep{Silk68} and nonlinearity \citep[e.g.,][]{Meiksin99,SE05,Jeong06,ESW07,Crocce08,Mat08}, $\kmax=0.3-0.5\ihMpc$ contains very little relevant information, but losing scales $k<0.3\ihMpc$ may be more damaging. \citet{Chang:2007xk} mention that a Nyquist sampled map with a pixel size of $\sim 18\hMpc$, and therefore $\kmax=\knyq=0.17\ihMpc$, is required for a BAO survey at $z\sim 1$, as the information on smaller scales is damped away due to nonlinearity. We want to probe this quantitatively: how much residual BAO information lies beyond $k\sim 0.17\ihMpc$? This question becomes especially pressing if reconstruction is anticipated, in which case we do not want the scales of interest for reconstruction to lie beyond the Nyquist frequency. 

We include the effect of the Nyquist frequency limit by introducing $\hat{W}$ as shown in equation (\ref{eq:StoN}) and (\ref{eq:window}), which effectively limits the integration range of the Fisher matrix calculation in \citet{Seo:2007ns}. We test the effect of including this limit as a function of $\kmax$.
The left panel of Figure \ref{fig:kmax} shows the Fisher matrix estimates of the errors on the the distance measurement $D_V$ as a function of $\kmax$ for different cases of $\nPt$ assuming a BAO feature at $z=1$ in redshift space; we assume a volume of $1\trihGpc$. With $\kmax \sim 0.2\ihMpc$ corresponding to a resolution of $16\hMpc$ (i.e., 23 arcmin at $z\sim 1$), compared to $\kmax >0.3\ihMpc$, the effect of introducing $\hat{W}$ is small when the instrumental noise is negligible (i.e., $\nPt \gg 10$). However, for a realistic survey with $\nPt < 1.5$ the error increases by $\sim 16\%$ relative to an estimate which includes small scales. This degradation in errors will be even more severe at high redshift, where the linear regime extends to smaller scales leaving even more information unobtainable at fixed telescope size.

The survey we consider is composed of a broad range of redshifts and therefore we observe an integrated effect of the Nyquist frequency limit over a range of redshift. 
 The middle panel of Figure \ref{fig:kmax} compares the FoM for our fiducial CRT configuration as a function of $\Lcyl$, i.e., angular resolution, with and without considering the Nyquist frequency limit, while the rest of telescope parameters are held fixed. The right panel shows the effect of the Nyquist frequency limit on the errors of $D_V$ at the individual redshift bins that comprise the fiducial survey. The effect is indeed larger at high redshift. Note that such an effect would have been much more severe than shown here at high redshift if not for the large survey area/volume available at high redshift; if we assume at $z=2$ the same volume as at $z=0.7$, the errors at $z=2$ will be $\sim 1.7$ times larger than shown here. Overall the effect of the Nyquist limit is small for a CRT array with a length larger than $100\sim 150$ meters (i.e., angular resolution better than $10-15$ arcmin at $z\sim 1$): $< 11\%$ in FoM with a Planck prior and even less with an addition of the DETF Stage II as priors. A smaller telescope will suffer more from the Nyquist frequency limit (magenta points). 

The second potential problem with limited resolution is related to the way the BAO feature is typically extracted from the power spectrum. To make the extraction as robust (systematics-free) as possible, the BAO feature is extracted by marginalizing over the broadband shape of the power spectrum. Poor angular resolution, which leads to a limited range of wavenumbers available, makes the correct estimate of the broad-band shape difficult, and this can bias the acoustic scale measurement, and therefore distance measurements, as well as increase the errors associated with the measurement. 

To test the second problem, we generate 61 random Gaussian density fields of $8\trihGpc$ (i.e., a total of $488 \trihGpc$) with $512^3$ density grids using a code~\citep{Sirko:2005uz} which produces initial conditions using second order Lagrangian perturbation theory. To account for nonlinear degradation of the BAO signal without evolving the density field, we smooth the input power spectrum based on the expected degradation at $z=1$ in real space. We Fourier-transform the ensuing distribution, compute the power spectrum, and conduct a $\chi^2$ analysis to measure the acoustic scale using Jackknife resampling. We follow the details of the $\chi^2$ analysis presented in \citet{Seo:2008yx}.

\begin{figure}
\plotone{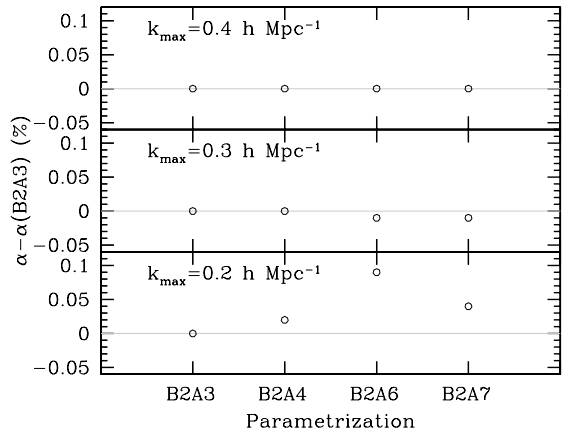}
\caption{The stability of the $\chi^2$ fitting (i.e., the differences between the best fits from different parametrizations) as a function of $\kmax$. The acoustic scale $\alpha$ is measured from the $\chi^2$ fitting to {\it one} Jackknife subsample (i.e., 480 $\trihGpc$) and the differences of $\alpha$ in the unit of the true acoustic scale are shown as a function of the parametrization for the broadband shape: for example, $B2A4$ means that we used a second order polynomial in $k$ for a scale-dependent bias $B(k)$ and a fourth order polynomial in $k$ for a smooth additive term $A(k)$ in \citet{Seo:2008yx}. Note that the fits using $\kmax >0.3\ihMpc$ display robust behaviour regardless of the parametrization, but when $\kmax \sim 0.2\ihMpc$ the best fit varies with parametrization.}\label{fig:Shiftonesim}
\end{figure}

In Figure \ref{fig:Shiftonesim}, we use {\it one} Jackknife subsample (i.e., 480 $\trihGpc$) and show the differences between the measured acoustic scales $\alpha$ as a function of the number of fitting parameters for the broadband shape. For example, $B2A4$ means that we used a second order polynomial in $k$ for a scale-dependent bias $B(k)$ and a fourth order polynomial in $k$ for a smooth additive term $A(k)$. A robust extraction of a BAO signal is characterized by a best fit $\alpha$ that is independent of the polynomial order for a range of reasonable choices of polynomials where ``reasonable choices'' means enough flexibility to account for the unknown broadband shape while not mimicking the BAO. From the figure, while $\kmax >0.3\ihMpc$ shows robust behaviour, when $\kmax \sim 0.2\ihMpc$ the best fit varies with parametrization: the standard deviation of the best fit values among these four parametrizations for the given subsample is 7 times larger for $\kmax=0.2 \ihMpc$ than for $\kmax=0.3 \ihMpc$. 

We find that the dispersion among {\it many} Jackknife subsamples (i.e, the precision associated with the measurement of the acoustic scale) is also sensitive to a choice of parametrization for $\kmax <0.2\ihMpc$. That is, for large values of $\kmax$, i.e., the higher resolution data, the choice of fitting formula is simple: all reasonable choices give the same result. However, in the case of poor resolution, the marginalizing process over the smooth component of the power spectrum becomes more challenging and the fit becomes highly dependent on the fitting formula. This sensitivity itself speaks to the peril of a low resolution survey.

In summary, angular resolution at the level of 20 arcmin (i.e., $\sim 16\hMpc$) at $z=1$ not only increases the noise of the survey but also limits the small wavenumber region where a residual BAO exists. The effect is more damaging at high redshift where the linear scales of interest correspond to smaller angular scales; still the impact on the DETF FoM is only ~$11\%$ for our fiducial survey.  Second, with a limited wavenumber region as a result of the poor angular resolution, it is difficult to construct a robust standard ruler test, and the extracted acoustic scale will be sensitive to the choice of fitting formula used to marginalize over the broadband shape. This means that we will need to know the broadband shape better, which diminishes the advantages of using the BAO as a standard ruler.

\begin{figure*}[t]
\epsscale{1.2}
\plotone{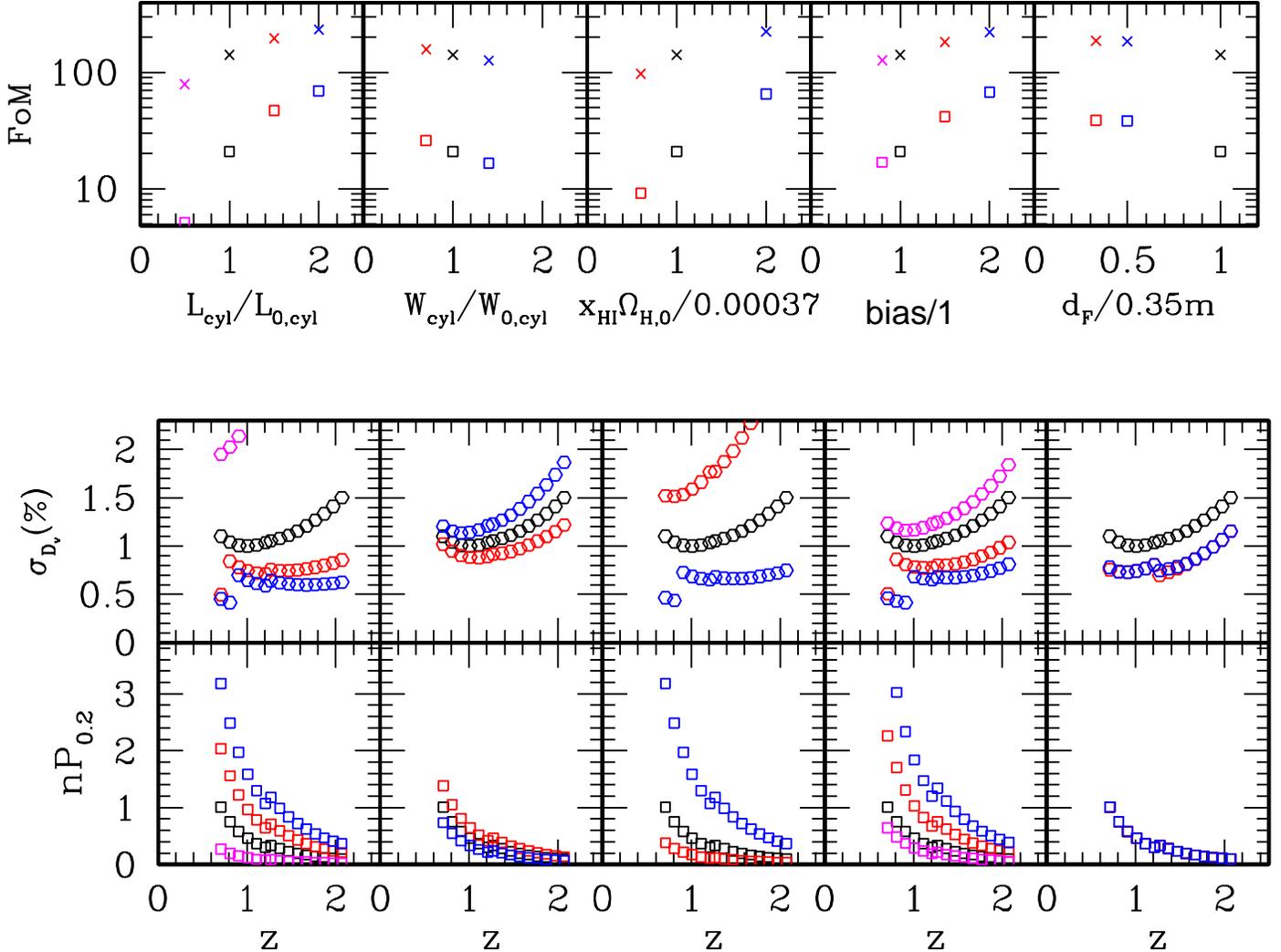}
\caption{FoM values as a function of various parameters (top panels). The square points are for FoM from CRT+Planck, and the cross points are for CRT+Planck+Stage II. The $x$-axis is scaled such that the fiducial model parameter presented in Table \ref{tab:fid} corresponds to $1$ in $x$-axis. The lower panels show $\nPt\equiv \Poh/(P_N+\Pshot)$ at $k=0.2\ihMpc$, which indicates the amount of signal relative to the noise at our scale of interest, and $\sigma_{D_V}$, the resulting error estimate on the distance scale $D_V$ at each redshift bin. The corresponding configuration between the top and the bottom panels is denoted with the same colored point. For example, black points correspond to our fiducial configuration. }\label{fig:figparam}
\end{figure*}

\section{FoM as a function of  telescope configurations}\label{sec:FoM}

In this section, we investigate the dependence of the final FoM values on various telescope configuration parameters. From equation (\ref{eq:StoN}), it is evident that, at a given $\Tsig$, the noise depends on the angular resolution, the integration time per pixel, the volume of the survey, and therefore on $\Lcyl$, $\Nyear/\Wcyl$, and the redshift range and $d_F$, respectively. The signal, which scales as $\Tsig$, will vary depending on our assumptions about $\OHI$, and clustering bias in $\Poh$. In Figure \ref{fig:figparam}, we show the dependence of the FoM from a CRT on each of these parameters while the rest of the parameters are fixed. We plot two values of FoM: first, when we combine the CRT with the Planck priors (square points), and second, when we add the DETF Stage II survey (cross points). The $x$-axis is scaled such that the fiducial model parameter presented in Table \ref{tab:fid} corresponds to unity. The lower panels show $\nPt\equiv \Poh/(P_N+\Pshot)$ at $k=0.2\ihMpc$, which indicates the amount of signal relative to the noise at our scale of interest, and $\sigma_{D_V}$, the resulting error estimate on the isotropic distance scale $D_V$ \citep{Eisenstein:2005su}, at each redshift bin for the corresponding configuration denoted with the same colored point in the top panels. The error $\sigma_{D_V}$ reflects both $\Vsur$ and $\nPt$ as well as redshift distortions.

With our fiducial survey (see Table \ref{tab:fid}), we find ${\rm FoM}=21$ without Stage II, and ${\rm FoM}=141$ including Stage II, which is $\sim 3$ times better than Stage II alone, and somewhat worse than Stage II + Stage III. From the bottom panels of the figure, one sees that our fiducial survey is noise-dominated at high redshift bins ($\nPt$ is less than one), mainly due to poor angular resolution. This results in a larger $\sigma_{D_V}$ in the high redshift bins, an effect somewhat offset by a larger volume per redshift bin at higher redshift.  

As expected, the FoM increases with increasing $\Lcyl$ due to increasing (i.e., better) angular resolution. Increasing $\Lcyl$ increases $\nPt$ and decreases $\sig_{D_V}$ over all redshift bins. While the improvement on $\sig_{D_V}$ is in general larger at high redshift, at the lowest redshift bins, $\nPt$ is now big enough to pass the reconstruction requirement of $\nPt >2$. The contribution from these lowest redshift bins is very important for improving the FoM
as $\Lcyl$ increases.

Increasing $\Wcyl$ decreases the amount of time that a celestial object will sit in the cylinder beam as it drifts across the cylinder width, i.e., the integration time, and therefore decreases the FoM. Note that decreasing $\Wcyl$ means increasing the number of cylinders and therefore increasing the number of feeds in the case of a compact array. Increasing and decreasing $\Nyear$ will affect the integration time in an opposite manner to $\Wcyl$, and therefore the two can be adjusted to balance the overall cost of the survey and the time required to complete it. 

For the survey we consider here, the receiver spacing $d_F$ determines the main lobe beam angle, therefore, the area of the sky covered. As $d_F$ reaches 0.17m, i.e., half of our fiducial spacing, the survey area increases to the whole sky, and no further improvement in FoM is obtained, as shown in the right panels of Figure \ref{fig:figparam}.  

The figure also shows that the neutral hydrogen fraction and the clustering bias of the neutral hydrogen are important uncertainties in estimating the performance of the CRT. Increasing the two parameters has two effects: first, the signal increases, therefore increasing the signal to noise. Second, some of the low redshift bins now can satisfy the $\nPt>2$ threshold for reconstruction, and therefore we can assume a smaller nonlinear degradation on the BAO feature for those redshifts. The clustering bias of the neutral hydrogen has been  investigated in several papers \citep{Meyer:2006gq,Wyithe:2009qj}. According to \citet{Marin}, large-scale bias of the neutral hydrogen is near unity at $z=1$, but increases to $\sim 1.3$ at $z\sim 2$. If we include this bias evolution in our fiducial model, we find a slightly larger FoM: 24 without Stage II and 152 with Stage II for our fiducial survey parameters.

\section{Maximization of FoM}\label{sec:Max}
In designing a future CRT array, we want to optimize the telescope configuration for the best performance (i.e., a largest FoM) while holding the cost fixed. In this section we describe an online calculator which performs this optimization.

For a given survey area, the figure of merit increases as the pixel noise and pixel size is reduced. However, reducing the pixel noise and pixel size increases the complexity of the CRT array. For example, while we have considered a square compact CRT array in the previous sections, we may want to sparsely locate cylinders to be more cost-effective, as the cost depends on the number of cylinders. The sensitivity per pixel, and therefore the noise per pixel, then will also depend on the packing factor, i.e., the ratio of the number of cylinders to the number of possible locations of the cylinders. 
\begin{figure}
\plotone{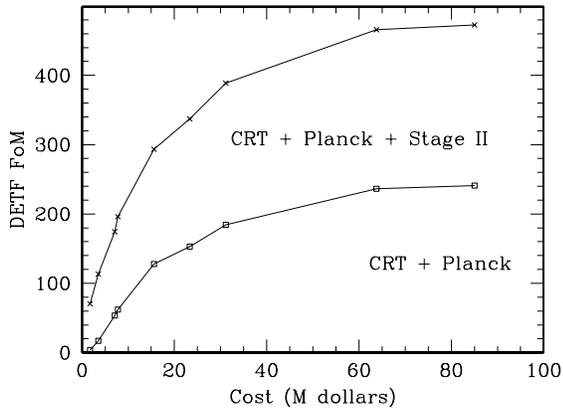}
\caption{Optimization of the CRT: the maximum DETF FoM at given cost. We assume a CRT at a latitude of 35 degrees.}\label{fig:opt}
\end{figure}
We can describe a cost function that parametrizes the complexity of the CRT array. It is not intended that these costs include everything that would arise in designing and building a large radio telescope, such as site preparation, non-recoverable engineering costs, overhead, contingency etc. These costs should only be used in trying to compare sets of design parameters. The cost function can be broken into three components; the cost of the digital electronics, the cost of the cylinder feed line, and the cost of the reflector surface.
\begin{eqnarray}
C_T&=& N_f N_c R_e + N_f N_c d_F R_f + N_f N_c d_f W^2_{\rm cyl} R_r
\end{eqnarray}
where $N_f$ is the number of feeds per cylinder, $N_c$ is the number of cylinders, $d_F$ is the feed spacing, and $\Wcyl$ is the width of the cylinders. In addition, $R_e$ is the cost per digital channel, $R_f$ is the feed line cost per unit length, and $R_r$ is the reflector cost per unit volume. These three cost rates can be scaled from the cost of existing prototypes. 

An optimization tool for the maximum FoM per given cost that takes into account a large number of CRT engineering parameters was developed and is available publicly at \url{http://astro.fnal.gov/21cm}.
 Figure \ref{fig:opt} shows the output of a design optimization for various CRT engineering parameters. 
Because fractional changes in array sensitivity and resolution drop off as the array cost increases, the corresponding figure of merit levels off as well. However, it should be noted that a FoM of 300 can be achieved with an array costing $\$15$ million according to our cost model.

\section{Conclusion}\label{sec:con}
In this paper we have considered a BAO survey with a ground-based 21cm intensity mapping with a CRT targeting a redshift range of 0.2-2.0. We have tested the feasibility of the survey while paying more careful attention to the effect of limited angular resolution. We summarize the results in this paper. 

First, we have tested the effect of angular resolution using the Fisher matrix calculations and Monte Carlos. The angular resolution at the level of 20 arcmin at $z\sim 1$ not only increases the instrumental noise beyond the signal but also notably limits the wavenumber range available for the standard ruler test. The latter effect further increases errors on the acoustic scale, while such effects on FoM are small for a CRT array with a size of 100-150 meters (i.e., an angular resolution of $10-14$ arcmin at $z\sim 1$). The limited wavenumber range also makes the marginalizing process over the smooth component of the power spectrum to isolate the BAO more challenging: the result highly depends on the fitting formula.  Therefore, in the case of poor resolution, the data are hard to fit reliably.

Second, we have investigated the dependence of the DETF FoM on various telescope parameters while assuming a simple compact CRT array. As expected, the FoM strongly depends on the telescope parameters as well as the neutral hydrogen fraction and the clustering bias.

Third, we have relaxed the assumption of the compact CRT array and have optimized the telescope configuration for the maximum FoM while holding the cost fixed.  We find that we can achieve FoM of 300 (with Planck and Stage II priors) for a reasonable cost.  The FoM/cost calculator and the optimization tool for a CRT is publicly available in \url{http://astro.fnal.gov/21cm}.
\acknowledgements
We would like to thank Patrick McDonald and Nickolay Y. Gnedin for extremely useful communications. H-JS, SD, JM, DM, AS, CS, and AV are supported by the U.S. Department of Energy under contract No. DE-AC02-07CH11359. 

\appendix
\section{21cm Intensity Power spectrum }
We derive the measured power spectrum in Fourier space, using the following definitions for discrete Fourier transform:
\begin{eqnarray}
&&\hdel_{\vec{k}}=\frac{1}{N}\sum_{x}e^{i\vec{k}\cdot\vec{x}}\delta_p(\vec{x})\\
&&\delta_p(\vec{x})=\sum_k \hdel_{\vec{k}} e^{-i\vec{k}\cdot\vec{x}} \\
&&\sum_x e^{i \vec{x}\cdot(\vec{k}-\vec{k'})}=N \delta^K_{\vec{k},\vec{k'}}\\
&&<\hdel_{\vec{k}}\hdel^*_{\vec{k'}}> V_\mu = P(\vec{k}) \delta^K_{\vec{k},\vec{k'}},
\end{eqnarray}
where $N$ is the total number of pixels in the map, $V_\mu$ is the total volume of the map under FFT, and $\delta^K$ is the Kronecker delta. Here $\delta_p$ is the fluctuations in power (i.e., Watts) measured for the pixel due to the large scale structure of HI ($p_S\delta_{HI}$) and instrument noise such that
\begin{eqnarray}
<\delta_p(\vec{x_p}) \delta_p(\vec{x_q})> &=&p^2_S(\xi_{HI})_{pq} + \sigma^2_{p_N} \delta^K_{pq},
\end{eqnarray}
where $\xi_{HI}$ is the two point correlation function of the neutral hydrogen density field. Parameters $p_S$ and $\sigma_{P_N}$ are the average power due to the 21cm line and the uncertainly associated with the average power $p_N$ due to the instrumental and sky noise, respectively, as defined in equation \ref{eq:pN} - \ref{eq:pS}.

Then,
\begin{eqnarray}
&&<\hdel_{\vec{k}}\hdel^*_{\vec{k'}}> = < \frac{1}{N} \sum_{x_p} e^{i\vec{k}\cdot\vec{x_p}} \delta_p(\vec{x_p})  \frac{1}{N} \sum_{x_q} e^{-i\vec{k'}\cdot\vec{x_q}} \delta_p(\vec{x_q})> \nonumber\\
&=& \frac{1}{N^2}\sum_{x_p}\sum_{x_q} <\delta_p(\vec{x_p}) \delta_p(\vec{x_q})>  e^{i\vec{k}\cdot\vec{x_p}}  e^{-i\vec{k'}\cdot\vec{x_q}} \nonumber \\
&=&\frac{1}{N^2}\sum_{x_p}\sum_{x_q} \left[p^2_S(\xi_{HI})_{pq} + \sigma^2_{p_N} \delta^K_{pq}\right] e^{i\vec{k}\cdot\vec{x_p}}  e^{-i\vec{k'}\cdot\vec{x_q}}\nonumber\\
&=&\frac{1}{N^2}\sum_{x_p}\sum_{x_q}\sum_{k''} \frac{p^2_S}{V_\mu} P_{HI}(\vec{k''}) e^{-i\vec{k''}\cdot(\vec{x_p}-\vec{x_q})} e^{i\vec{k}\cdot\vec{x_p}}  e^{-i\vec{k'}\cdot\vec{x_q}} \nonumber\\
&&+ \frac{1}{N^2}\sum_{x_p} \sigma^2_{p_N} e^{i(\vec{k}-\vec{k'})\cdot\vec{x_p}} \nonumber\\ 
&=&\frac{1}{N^2}\sum_{k''} \frac{p^2_S}{V_\mu}  P_{HI}(\vec{k''}) \sum_{x_p}\sum_{x_q}  e^{i(\vec{k}-\vec{k''})\cdot\vec{x_p}} e^{i(\vec{k''}-\vec{k'})\cdot\vec{x_q}} \nonumber\\
&& + \frac{1}{N}\sigma^2_{p_N} \delta^K_{\vec{k},\vec{k'}} \nonumber\\
&=& \sum_{k''} \frac{p^2_S}{V_\mu}  P_{HI}(\vec{k''})  \delta^K_{\vec{k},\vec{k''}}\delta^K_{\vec{k''},\vec{k'}} +  \frac{1}{N}\sigma^2_{p_N}\delta^K_{\vec{k},\vec{k'}} \nonumber \\
&=& \frac{p^2_S}{V_\mu}  P_{HI}(\vec{k})  \delta^K_{\vec{k},\vec{k'}} +  \frac{1}{N}\sigma^2_{p_N} \delta^K_{\vec{k},\vec{k'}},
\end{eqnarray}
where $P_{HI}(\vec{k})$ is the power spectrum in $h^{-3}{\rm Mpc}^3$ due to the large scale structure of the neutral hydrogen, i.e., our signal.
The measured power spectrum of the 21cm intensity $\hat{P}(\vec{k})$ is then, 
\begin{eqnarray}
<\hdel_{\vec{k}}\hdel^*_{\vec{k'}}> V_\mu &=& \hat{P}(\vec{k}) \delta^K_{\vec{k},\vec{k'}} = p^2_S P_{HI}(\vec{k})  \delta^K_{\vec{k},\vec{k'}} +  \frac{V_\mu}{N}\sigma^2_{p_N} \delta^K_{\vec{k},\vec{k'}}\nonumber \\
\hat{P}(\vec{k})&=&p^2_S P_{HI}(\vec{k}) + V_R \sigma^2_{p_N},
\end{eqnarray}
where $V_R$ is the volume of a pixel.



\begin{thebibliography}{14}
\expandafter\ifx\csname natexlab\endcsname\relax\def\natexlab#1{#1}\fi

\bibitem[Albrecht et al.(2006)]{Albrecht:2006um} Albrecht, A., et al.\ 
2006, arXiv:astro-ph/0609591 

\bibitem[Ansari et al.(2008)]{Ansari:2008yw} Ansari, R., Le Goff, 
J.~-., Magneville, C., Moniez, M., Palanque-Delabrouille, N., Rich, J., 
Ruhlmann-Kleider, V., \& Y{\`e}che, C.\ 2008, arXiv:0807.3614 


\bibitem[{Barkana \& Loeb(2007)}]{Barkana07}
Barkana, R. \& Loeb, A. 2007, Rept. Prog. Phys., 70, 627

\bibitem[Blake \& Glazebrook(2003)]{Blake03} Blake, C., \&
Glazebrook, K.\ 2003, \apj, 594, 665


\bibitem[{Chang {et~al.}(2008)Chang, Pen, Peterson, \& McDonald}]{Chang:2007xk}
Chang, T.-C., Pen, U.-L., Peterson, J.~B., \& McDonald, P. 2008, Phys. Rev.
  Lett., 100, 091303

\bibitem[Crocce \& Scoccimarro(2008)]{Crocce08} Crocce, M., \& Scoccimarro, R.\ 2008, \prd, 77, 0235
33

\bibitem[{Di Matteo {et~al.}(2002)}]{DiMatteo:2001gg}
Di Matteo, T., Perna, R., Abel, T., \& Rees, M. 2002, Astrophys. J., 564, 576


\bibitem[Eisenstein \& Hu(1998)]{EH98} Eisenstein, D.~J., \&
Hu, W.\ 1998, \apj, 496, 605

\bibitem[Eisenstein et al.(1998)]{EHT98} Eisenstein, D.~J.,
Hu, W., \& Tegmark, M.\ 1998, \apjl, 504, L57

\bibitem[Eisenstein(2003)]{Eisen03}
        Eisenstein, D.J., 2003, in ASP Conference Series, volume 280, Next Generation Wide Field Multi-Object Spectroscopy,
        ed. M.J.I. Brown \& A. Dey (ASP: San Francisco) pp. 35-43;
        astro-ph/0301623

\bibitem[{Eisenstein {et~al.}(2005)}]{Eisenstein:2005su}
Eisenstein, D.~J. {et~al.} 2005, Astrophys. J., 633, 560

\bibitem[{Eisenstein {et~al.}(2007)Eisenstein, Seo, Sirko, \& Spergel}]{Eisenstein:2006nk}
Eisenstein, D.~J., Seo, H.-J., Sirko, E., \& Spergel, D. 2007, Astrophys. J.,
  664, 675

\bibitem[Eisenstein et al.(2007)]{ESW07} Eisenstein, D.~J.,
Seo, H.-J., \& White, M.\ 2007, \apj, 664, 660

\bibitem[Holtzman(1989)]{Holtzman89} Holtzman, J.~A.\ 1989, \apjs, 
71, 1 

\bibitem[Hu \& Sugiyama(1996)]{Hu96} Hu, W., \& Sugiyama, N.\ 1996, \apj, 471, 542 

\bibitem[Hu 
\& White(1996)]{HW96} Hu, W., \& White, M.\ 1996, \apj, 471, 30 

\bibitem[Hu \& Haiman(2003)]{Hu03} Hu, W., \& Haiman, Z.\
2003, \prd, 68, 063004

\bibitem[Jeong \& Komatsu(2006)]{Jeong06} Jeong, D., \&
Komatsu, E.\ 2006, \apj, 651, 619

\bibitem[Kaiser(1987)]{Kaiser87} Kaiser, N.\ 1987, \mnras, 227,
1

\bibitem[Loeb \& Wyithe(2008)]{LW08} Loeb, A., \& Wyithe, J.~S.~B.\ 2008, Physical Review Letters, 100, 161301 


\bibitem[Linder(2003)]{Linder03} Linder, E.~V.\ 2003, \prd, 68, 
083504 

\bibitem[{Mao {et~al.}(2008)Mao, Tegmark, McQuinn, Zaldarriaga, \& Zahn}]{Mao:2008ug}
Mao, Y., Tegmark, M., McQuinn, M., Zaldarriaga, M., \& Zahn, O. 2008, Phys.
  Rev., D78, 023529

\bibitem[Marin et al.(2009)]{Marin} Marin, F., Gnedin, N.~Y., 
Seo, H.-J., \& Vallinotto, A.\ 2009, arXiv:0911.0041 

\bibitem[Matsubara(2008)]{Mat08} Matsubara, T.\ 2008, \prd,
77, 063530

\bibitem[Meiksin, White, \& Peacock(1999)]{Meiksin99}
        Meiksin, A., White, M., \& Peacock, J.~A.\ 1999, \mnras, 304, 851

\bibitem[{Meyer {et~al.}(2007)Meyer, Zwaan, Webster, Brown, \& Staveley-Smith}]{Meyer:2006gq}
Meyer, M.~J., Zwaan, M.~A., Webster, R.~L., Brown, M. J.~I., \& Staveley-Smith,
  L. 2007, Astrophys. J., 654, 702

\bibitem[{Morales {et~al.}(2006)}]{Morales:2005qk}
Morales, M., Bowman, J.~D., \& Hewitt, J.~N. 2006, Astrophys. J., 648, 767

\bibitem[Parkinson et al.(2007)]{Parkinson07} Parkinson, D., Blake, 
C., Kunz, M., Bassett, B.~A., Nichol, R.~C., 
\& Glazebrook, K.\ 2007, \mnras, 377, 185 

\bibitem[Parkinson et al.(2009)]{Parkinson09} Parkinson, D., Kunz, 
M., Liddle, A.~R., Bassett, B.~A., Nichol, R.~C., 
\& Vardanyan, M.\ 2009, arXiv:0905.3410 

\bibitem[Peebles 
\& Yu(1970)]{Peebles70} Peebles, P.~J.~E., \& Yu, J.~T.\ 1970, \apj, 162, 815 

\bibitem[Pritchard 
\& Loeb(2008)]{Prit08} Pritchard, J.~R., \& Loeb, A.\ 2008, \prd, 78, 103511 

\bibitem[{{Prochaska} {et~al.}(2005){Prochaska}, {Herbert-Fort}, \&
  {Wolfe}}]{igm:phw05}
{Prochaska}, J.~X., {Herbert-Fort}, S., \& {Wolfe}, A.~M. 2005, \apj, 635, 123

\bibitem[{{Rao} {et~al.}(2006){Rao}, {Turnshek}, \& {Nestor}}]{igm:rtn06}
{Rao}, S.~M., {Turnshek}, D.~A., \& {Nestor}, D.~B. 2006, \apj, 636, 610


\bibitem[Rybicki \& Lightman(1985)]{Ryb} Rybicki, G.~B., \& Lightman, A.~P.\ 1985, Radiative processes in astrophysics (Wiley: NY)

\bibitem[Seo \& Eisenstein(2003)]{SE03} Seo, H.-J., \&
Eisenstein, D.~J.\ 2003, \apj, 598, 720

\bibitem[Seo \& Eisenstein(2005)]{SE05} Seo, H.-J., \&
Eisenstein, D.~J.\ 2005, \apj, 633, 575


\bibitem[{Seo \& Eisenstein(2007)}]{Seo:2007ns}
Seo, H.-J. \& Eisenstein, D.~J. 2007, Astrophys. J., 665, 14

\bibitem[{Seo {et~al.}(2008)Seo, Siegel, Eisenstein, \& White}]{Seo:2008yx}
Seo, H.-J., Siegel, E.~R., Eisenstein, D.~J., \& White, M. 2008, Astrophys. J.,
  636, 13

\bibitem[Silk(1968)]{Silk68} Silk, J.\ 1968, \apj, 151, 459

\bibitem[{Sirko(2005)}]{Sirko:2005uz}
Sirko, E. 2005, Astrophys. J., 634, 728

\bibitem[{Spergel {et~al.}(2003)}]{Spergel:2003cb}
Spergel, D.~N. {et~al.} 2003, Astrophys. J. Suppl., 148, 175

\bibitem[{Stebbins(2010)}]{Stebbins:2010}
Stebbins, A. in preparation

\bibitem[Visbal et al.(2009)]{Visbal09} Visbal, E., Loeb, A., 
\& Wyithe, S.\ 2009, Journal of Cosmology and Astro-Particle Physics, 10, 30 


\bibitem[{Wang {et~al.}(2005)}]{Wang:2005zj}
Wang, X.-M., Tegmark, M., Santos, M., \& Knox, L. 2006, Astrophys. J., 650, 529

\bibitem[Wyithe \& Loeb(2008)]{WL08} Wyithe, J.~S.~B., \& Loeb, A.\ 2008, \mnras, 383, 606 

\bibitem[Wyithe et al.(2008)]{WLal08} Wyithe, J.~S.~B., Loeb, 
A., \& Geil, P.~M.\ 2008, \mnras, 383, 1195 

\bibitem[{Wyithe {et~al.}(2009)Wyithe, Brown, Zwaan, \& Meyer}]{Wyithe:2009qj}
Wyithe, S., Brown, M. J.~I., Zwaan, M.~A., \& Meyer, M.~J. 2009

\bibitem[Wyithe \& Loeb(2009)]{WL09} Wyithe, J.~S.~B., \& Loeb, A.\ 2009, \mnras, 397, 1926 


\bibitem[{Zwaan {et~al.}(2005)}]{Zwaan:2005cz}
Zwaan, M.A., Meyer, M.J., Staveley-Smith, L., \& Webster, R.L. 2005, MNRAS, 239, L30

\bibitem[{Zaldarriaga {et~al.}(2004)}]{Zaldarriaga:2003du}
Zaldarriaga, M., Furlanetto, S.~R., \& Hernquist, Lars 2005, Astrophys. J., 608, 622

\end{thebibliography}

\end{document}